\documentclass[10pt,letterpaper]{article}
\usepackage{opex3}
\usepackage{cite}
\usepackage{graphicx}

\begin{document}

\title{Super-resolution image transfer by a vortex-like metamaterial}
\author{Hui Yuan Dong,$^{1,2}$~Jin Wang,$^1$~Kin Hung Fung,$^3$~and Tie Jun Cui$^{4,*}$}

\address{
\small $^1$Department of Physics, Southeast University, Nanjing 211189, China \\
\small $^2$School of Science, Nanjing University of Posts and Telecommunications, Nanjing 210003, China\\
\small $^3$Department of Applied Physics, The Hong Kong Polytechnic University, Hong Kong, China\\
\small $^4$State Key Laboratory of Millimeter Waves, Department of Radio Engineering, Southeast University, Nanjing 210096, China}
\email{*tjcui@seu.edu.cn}

\begin{abstract}
We propose a vortex-like metamaterial device that is capable of transferring image along a spiral route without losing subwavelength information of the image. The super-resolution image can be guided and magnified at the same time with one single design. Our design may provide insights in manipulating super-resolution image in a more flexible manner. Examples are given and illustrated with numerical simulations.
\end{abstract}


\section{Introduction}

Recently, a great deal of attention has been devoted to metamaterials because such materials can be artificially engineered to have remarkable properties, such as negative index of refraction\cite{shelby}, invisibility cloaking\cite{liu}, and superimaging/hyperimaging\cite{pendry,fang,taubner,smoly,liu3,salan,belov,jacob,liu2,rho,kild,wang,cast,lu,kawata,zhao}, which are difficult or impossible to achieve with natural materials. Initiated by Pendry's seminal concept of the perfect lens that could ideally image a perfect copy of a source\cite{pendry}, a great variety of superlens\cite{fang,taubner,smoly,liu3,salan,belov} were proposed to achieve sub-diffraction-limited resolution. Fang $et~al.$\cite{fang} demonstrated experimentally the feasibility of subwavelength imaging by enhancing evanescent waves through a slab of silver in the optical frequency range. However, such a silver superlens can only work for the near field, which makes the image difficult to be processed or brought to a focus by conventional optics.  To solve this problem, the hyperlens based on metamaterial crystals was proposed\cite{jacob} and experimentally realized\cite{liu2,rho}, which has the ability to form a magnified image of a subwavelength object in the far field.

In general, the hyperlens is a hollow core cylinder or half-cylinder made of anisotropic materials, with the two permittivity tensor components in the radial direction, $\epsilon_{r}$, and the tangential direction, $\epsilon_{\theta}$, being opposite in sign, resulting in a hyperbolic dispersion relation for TM-polarized waves. The simplest anisotropic metamaterials can be constructed by the deposition of alternating metal-dielectric multilayers. By bending the flat layers into concentrically curved layers, we could compress progressively the tangential wave-vectors as the waves travel along the radial direction. Therefore, a magnified image carrying information about the detailed structure of the object could propagate into the far field\cite{jacob}. However, owing to its use of curved surfaces, such a hyperlens may not be convenient for practical applications.

In this work, we present a novel hyperlens made of vortex-like metamaterials that is capable of manipulating subwavelength images along a spiral route. In one single design, super-resolution image transfer through a curved route of arbitrary angle, long-range image transfer, plane-to-plane imaging and image magnification can be achieved. Such a multi-functional design may contribute to potential applications such as biomedical imaging and nanolithography.
\begin{figure}[htb]
\centerline{\includegraphics[width=10.5cm]{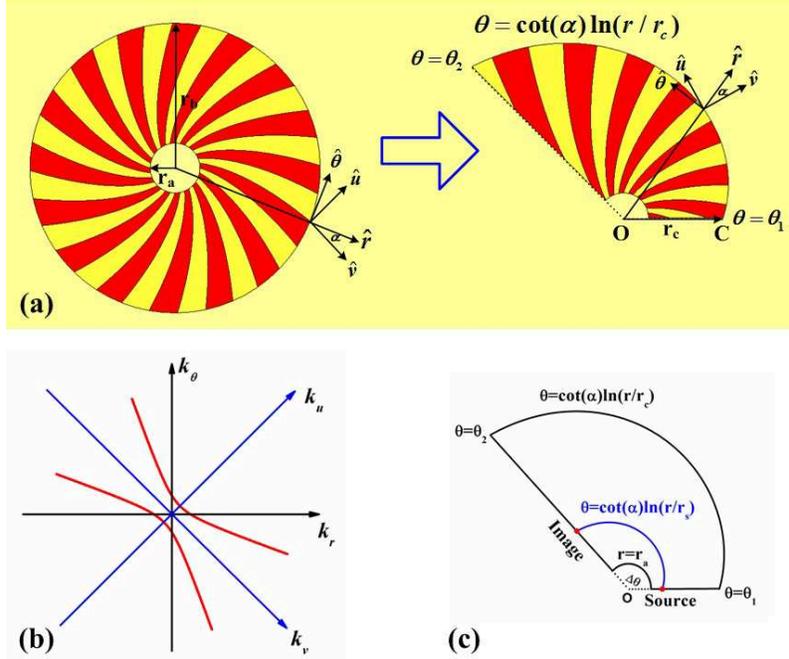}}
\caption{(a) The geometry of the proposed vortex-like layered hyperlens in the $r$-$\theta$ plane, the $\hat{u}$ and $\hat{v}$ directions are two principal axes obtained by rotating an angle $\alpha$ from the $\hat{\theta}$ and $\hat{r}$. (b) Typical hyperbolic equifrequency contours (EFCs) pertaining to the dispersion relation in Eq. (\ref{efc}) for $\epsilon_{u}<0$ and $\epsilon_{v}>0$ in the cylindrical ($k_{r},k_{\theta}$) and rotated $(k_{u},k_{v})$ spectral reference systems. (c) Sketch of the idea of image rotating setup.  The point source is placed at the input plane of $\theta=\theta_{1}$, and the distance $r_{s}$ away from the origin $O$ of the hyperlens.}
\end{figure}

\section{Principle and design}

We begin with a vortex-like $N$-layered structure shown in the left panel of Fig. 1(a) ($r_a$ and $r_b$ are, respectively, the inner and outer radii of the metamaterial device) with alternating layers of dielectric and plasmonic materials whose permittivities are $\epsilon_{d}$ and $\epsilon_{m}$, respectively. For the shape of each layer, we have the formula in the cylindrical coordinates ($r$,$\theta$,$z$) [the $\hat{z}$ is homogenous]\cite{wang,chen}
\begin{equation}
\theta=\beta+\tan(\alpha)\ln\frac{r_{a}}{r}
\end{equation}
with the starting point in the inner circle, $r=r_a$ and $\theta=\beta$. Here, $\alpha$ is an oblique angle, and $\beta=0$, $2\pi/N$, $4\pi/N$, ..., $2(N-1)\pi/N$. Starting with these $N$ points, we can produce $N$ curves that inside the metamaterials ($r_{a}<r<r_{b}$) into $N$ fan-shaped parts shown in Fig. 1(a). We assume that the magnetic field is perpendicular to $r$-$\theta$ plane (TM polarization, magnetic field in the $\hat{z}$ direction) and the time harmonic factor is $\exp(-i\omega t)$.

Next, let us cut the layered system along two lines of $\theta=\theta_{1}$ and $\theta=\theta_{2}$, which are taken respectively as the input (source) plane and output (image) plane, and two other fixed curves $r=r_a$ and $\theta=\cot(\alpha)\ln(r/r_{c})$ (with $r_c$ the distance between the origin $O$ and a fixed point $C$, and $r_{a}<r_{c}<r_{b}$), which are considered as the interfaces of the hyperlens shown in the right panel of Fig. 1(a). For the simplicity and generality of illustration, we consider the normal direction of the alternating layers as the $\hat{u}$ direction, by rotating at a fixed angle, $\alpha$, from the $\hat{\theta}$ direction and the $\hat{v}$ direction as another principal axis perpendicular to the $\hat{u}$ direction. In the long-wavelength limit, the optical properties of the layered system can be represented by an anisotropic effective medium in the principal coordinate system ($u,v$),
\begin{equation}
\bar{\epsilon}_{u,v}=
\epsilon_{0} \left [
\begin{tabular}{ccc}
$\epsilon_{u}$ & $0$  \\
$0$ & $\epsilon_{v}$
\end{tabular}
 \right ], \label{perm}
\end{equation}
where $\epsilon_{u}=2\epsilon_{d}\epsilon_{m}/(\epsilon_{d}+\epsilon_{m})$ and $\epsilon_{v}=(\epsilon_{d}+\epsilon_{m})/2$.

The kinematical properties of wave propagation in such an effective medium could be understood by looking at the equifrequency contours (EFCs) of the following dispersion relation\cite{cast},
\begin{equation}
\frac{k_{u}^2}{\epsilon_{v}}+\frac{k_{v}^2}{\epsilon_{u}}=k_{0}^{2}, \label{efc}
\end{equation}
with $k_{0}=\omega\sqrt{\epsilon_{0}\mu_{0}}=2\pi/\lambda$ denoting the vacuum wave number (and $\lambda$ the corresponding wavelength), and $k_{u}$ and $k_{v}$ referring to wave vector components of the principal coordinate system $(u,v)$ illustrated by Fig. 1(b). The equation may be either elliptic or hyperbolic, depending on the signs of the permittivity tensor elements. When sign$(\epsilon_{u})\neq$ sign$(\epsilon_{v})$, the EFCs represents a hyperbola. More specifically, when $\epsilon_{u}<0$, $\epsilon_{v}>0$ and $\epsilon_{v}\rightarrow 0$, the hyperbolic EFCs tends to be very flat along the $v$ direction, which allows high spatial frequency components of any field propagate close to the $u$ direction in a raylike fashion, thereby in our proposed hyperlens subwavelength details can be transported along a particular direction $\theta=\cot(\alpha)\ln(r/r_{s})$ from source point to image point indicated by Fig. 1(c), when the point source is placed at the input plane of $\theta=\theta_{1}=0^{\circ}$, and the distance $r_{s}$ away from the origin of the hyperlens.
\begin{figure}[htb]
\centerline{\includegraphics[width=8.5cm]{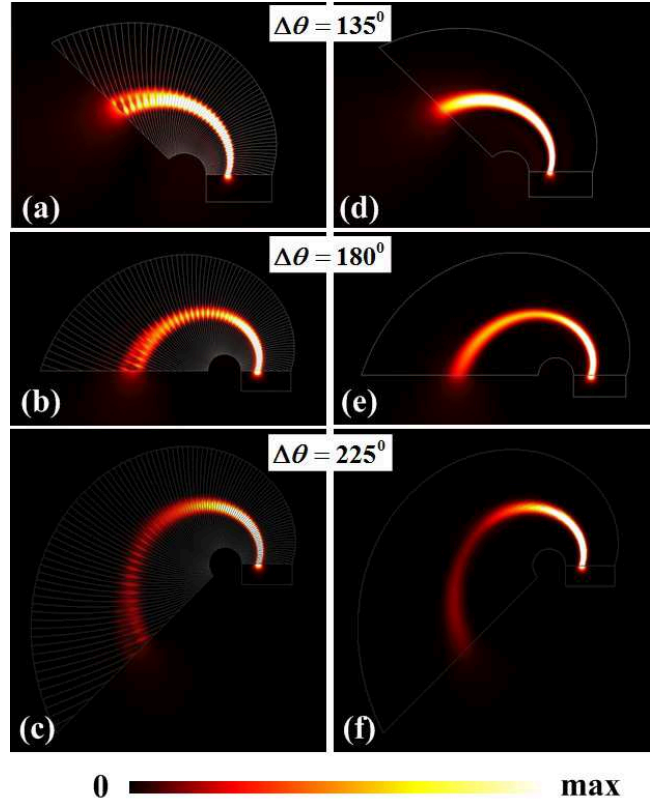}}
\caption{The distribution of magnetic energy density of a 2D transverse magnetic (TM) excitation for a 2D layered hyperlens [(a)-(c)] and the corresponding effective anisotropic medium [(d)-(f)], respectively. Three cases of different angles of rotation $\Delta\theta=135^{\circ}$, $180^{\circ}$ and $225^{\circ}$ are considered.  The point source is placed at the input plane of $\theta=\theta_{1}=0^{\circ}$, and the distance $r_{s}=0.2\lambda$ away from the origin of the hyperlens. The oblique angle is taken as $\alpha=18^{\circ}$ and $r_{a}=0.1\lambda$, $r_{c}=0.4\lambda$.}
\end{figure}

\section{Simulation and discussion}

We consider a practical and lossy configuration featuring a point source placed at the input plane $\theta=\theta_{1}=0^{\circ}$ and a distance $r_{s}=0.2\lambda$ away from the origin of the hyperlens, constitutive parameters $\epsilon_{m}=-4+i0.25$, $\epsilon_{d}=4.3$. The mask near the input plane is used to block the input light in our simulations. Figures 2(a)-2(c) show the image of a point source through three layered structures for different rotation of image with $\Delta\theta=\theta_{2}-\theta_{1}=135^{\circ}, 180^{\circ}$ and $225^{\circ}$. As we expect, light rays are bent along a fixed curve $\theta=\cot(\alpha)\ln(r/r_{s})$ and form a resolvable image at a flat output plane $\theta=\theta_{2}$, which is in agreement with the fact that subwavelength details of the source are effectively transported with small distortion. It is observed that worse localization and weaker intensity of images are accompanied by larger angle rotation of image. To give us intuitive guidance, we treat the real layered structure as an effective medium given by Eq. (\ref{perm}) and achieve very consistent results [see Figs. 2(d)-2(f)].

\begin{figure}[htb]
\centerline{\includegraphics[width=11.5cm]{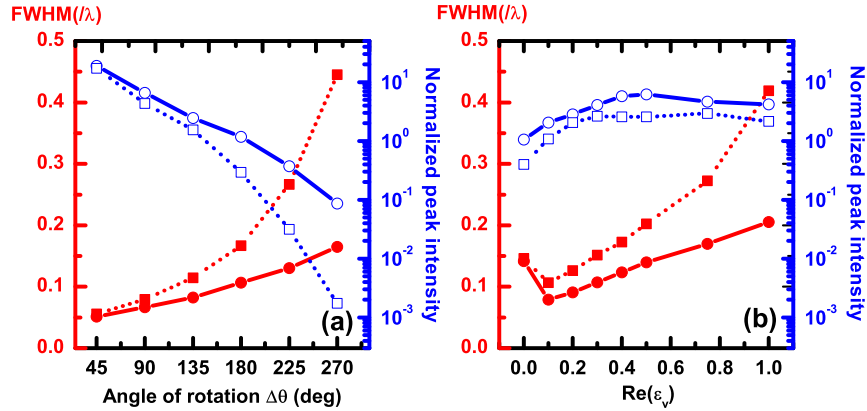}}
\caption{FWHM (full markers, left axis) and peak intensity (empty markers, right axis) at the output plane, for two cases of the oblique angle $\alpha=18^{\circ}$ (solid lines), and $\alpha=27^{\circ}$ (dotted lines), as a function of (a) angle of rotation $\Delta\theta$, and (b) effective permittivity Re$(\epsilon_{v})$.}
\end{figure}

For quantitative assessments, we consider two typical figures of merit: the $full$-$width$-$at$-$half$-$maximum$ (FWHM) and the (normalized) peak intensity at the image plane $\theta=\theta_{2}$. Figure 3(a) gives the FWHM and (normalized) peak intensity as a function of $\theta$ for two cases of different oblique angle $\alpha=18^{\circ}$ and $27^{\circ}$. Consistently with the the visual impression from Fig. 2, both of the observables deteriorate. More specifically, over the range $45^{\circ}<\theta<270^{\circ}$ in the case of $\alpha=18^{\circ}$, the FWHM increases from $\sim0.05\lambda$ to $\sim0.16\lambda$, while the (normalized) peak intensity decreases from $\sim18$ to $\sim0.09$. However, larger oblique angle [i.e. $\alpha=27^{\circ}$ shown by dotted lines in Fig. 3(a)] results in a larger degradation of the image resolution and intensity. As we tune the permittivity $\epsilon_{d}$ of dielectric layer, and the real part of effective parameter $\epsilon_{v}$ deviates away from zero, the image quality will be reduced rapidly, while the peak intensity varies slowly as seen from Fig. 3(b). However, just for Re$(\epsilon_{v})=0$, over-amplification of some evanescent waves can not be eliminated completely, it may deteriorate image quality to some extent \cite{wang}.

\begin{figure}[htb]
\centerline{\includegraphics[width=8.5cm]{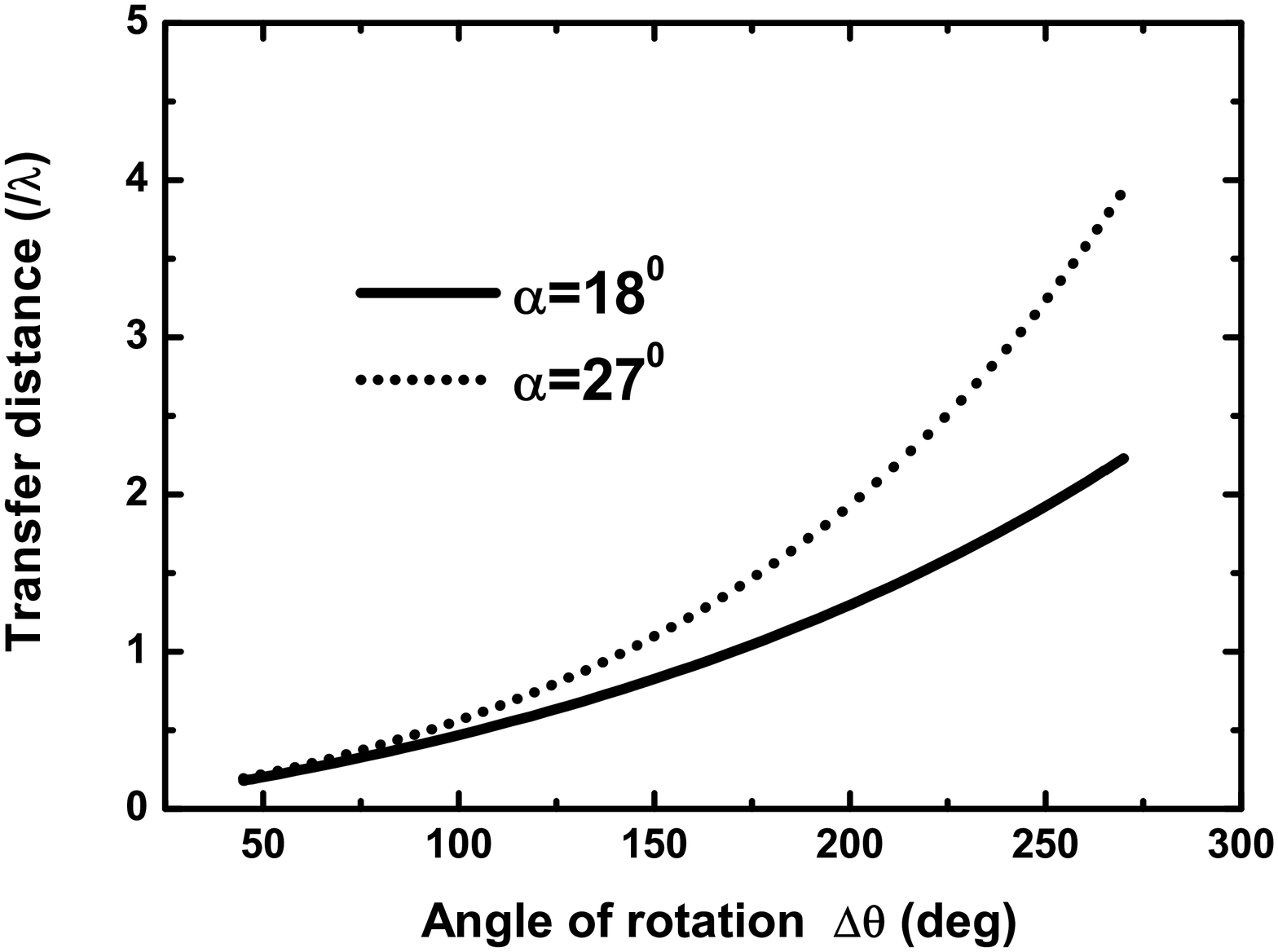}}
\caption{Image transfer distance as a function of angle of rotation $\Delta\theta$ when $\alpha$ are taken as $18^{\circ}$ (solid lines), and $27^{\circ}$ (dotted lines), respectively. }
\end{figure}

On the other hand, we show in Fig. 4 that subwavelength image could be transferred over several wavelengths for larger $\alpha$ and $\Delta\theta$. For instance, when $\alpha=27^{\circ}$ and $\Delta\theta=270^{\circ}$, image transfer distance of nearly four wavelengths could be attained, which follows the formula $s=\int_{\theta_{1}}^{\theta_{2}} r d\theta=r_s \cot(\alpha) (\exp[\Delta\theta/\cot(\alpha)]-1)$. Therefore such a hyperlens could act as an optical component that makes long-distance transfer of subwavelength image between unparallel input and output planes, simultaneously with an arbitrary angle of rotation, $\Delta\theta=\theta_{2}-\theta_{1}$ with respect to the origin of the hyperlens.

\begin{figure}[htb]
\centerline{\includegraphics[width=7.5cm]{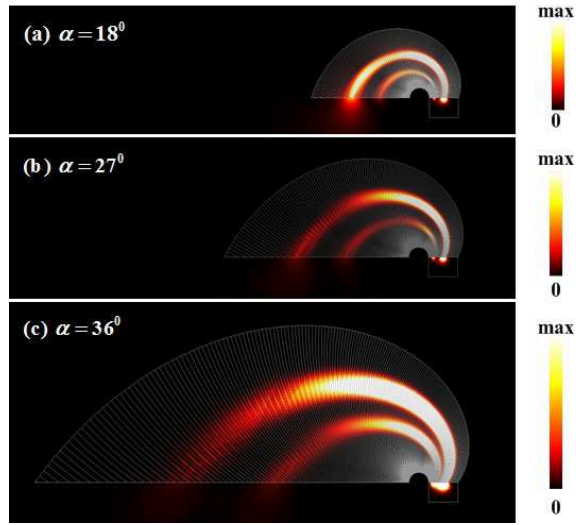}}
\caption{The distribution of magnetic energy density for the layered hyperlens with the oblique angle $\alpha=18^{\circ}, 27^{\circ}$ and $36^{\circ}$. Two point sources at the input plane $\theta=\theta_{1}=0^{\circ}$ are separated by $\Delta r=0.1\lambda$, where the ratio of the source amplitudes is $1:3$.}
\end{figure}
\begin{figure}[htb]
\centerline{\includegraphics[width=9.5cm]{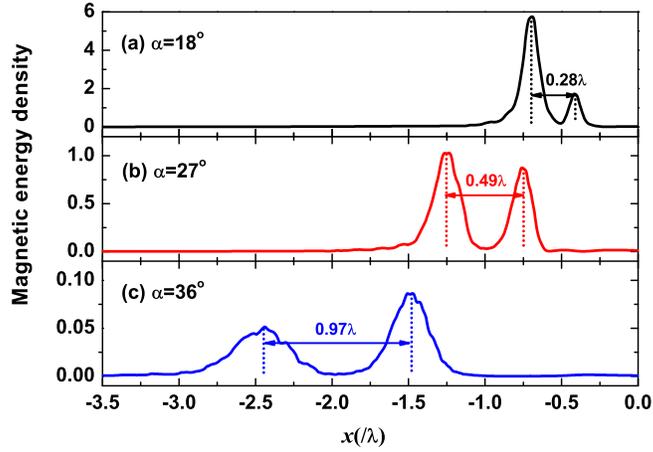}}
\caption{Line scans at the image plane $\theta=\theta_{2}=180^{\circ}$ of the magnetic energy density for different cases of oblique angle $\alpha=18^{\circ}, 27^{\circ}$ and $36^{\circ}$. Other parameters are the same as Fig. 5.}
\end{figure}

Moreover, for our proposed hyperlens, if we put two sources on the input plane with their distance $\Delta r$, their images will be transferred to the output plane by means of two curved rays [along the direction $\theta=\cot(\alpha)\ln(r/r_{s})$], and the image separation will appear to be $\Delta r'=\Delta r \exp[\Delta\theta/\cot(\alpha)]$ on that plane. Since $\Delta\theta/\cot(\alpha)$ could be much greater than zero, images generated on the output plane could be far enough apart to be resolved by a conventional optical device on the output plane. We represent in Fig. 5 the functionality of the image magnification through the proposed hyperlens for three configurations of different oblique angle $\alpha$ with a fixed angle of rotation $\Delta\theta=180^{\circ}$. Two point sources at the input plane separated by $\Delta r=0.1\lambda$ have been well imaged at the output plane. The magnification could even approach to be near $10$ times shown in Fig. 5(c). The line scans at the output plane of magnetic energy density have been illustrated in Fig. 6, which clearly indicate planar objects with deep subwavelength features can be projected and magnified to wavelength scale planar image. Unfortunately, the special dispersion of such a vortex-like metamaterial and its geometry provide only 1D magnification\cite{salan}. Nevertheless, compared to the ``oblique cut" geometry in Ref. \cite{salan}, the proposed vortex-like structure can transfer subwavelength images along a spiral route of any rotating angle, which provides a more flexible way to manipulate super-resolution image. It is noticed that the images display different magnitudes of intensity due to the variant length of light path, which leads to different losses. This effect can, in principle, be compensated through a nonuniform illumination of the input plane\cite{salan}, or by designing a compensating lens to ensure the lengths of ray traces from input plane to the output plane are the same\cite{zhao}.

We remark that practical fabrication of such a vortex-like metamaterial could be challenging. However, it is found that complicated ultrafine nanostructures could be fabricated with single-nanometer precision in both feature size and location. Therefore, it is expected that several advanced fabrications may enable us to achieve such a vortex-like structure\cite{won,zhang}.

\section{Conclusion}

In conclusion, we demonstrated that a vortex-like metamaterials is capable of manipulating subwavelength images along a spiral route of arbitrary turning angle. Using the same design, long-range image transfer, plane-to-plane imaging and image magnification can also be achieved.

\section*{Acknowledgments}

This work was supported in part by the National Science Foundation of China under Grant Nos. 11204036, 60990320, 60990324, 61138001, and 60921063, in part by the National High Tech (863) Projects under Grant Nos. 2011AA010202 and 2012AA030402, in part by the 111 Project under Grant No. 111-2-05, and in part by the Campus Funding No. NY210050 from Nanjing University of Posts and Telecommunications.

\end{document}